\documentclass[10pt,twocolumn,letterpaper]{article}

\usepackage[margin=0.75in]{geometry}
\usepackage{epsfig}
\usepackage{times}
\usepackage{amsmath}
\usepackage{amssymb}
\usepackage{amsfonts}
\usepackage{url}
\usepackage{graphicx}
\usepackage[utf8]{inputenc}
\usepackage{float}
\usepackage{bm}
\usepackage{color}
\usepackage{tabularx}
\usepackage{mathtools}
\usepackage{multirow}
\usepackage{siunitx}
\usepackage{comment}
\usepackage{placeins}

\usepackage[breaklinks=true,bookmarks=false]{hyperref}

\DeclareMathOperator{\rect}{rect}
\DeclareMathOperator{\erf}{erf}
\DeclareMathOperator{\atan}{atan}
\newcommand{\pixel}{\mathbf{p}}
\renewcommand{\eqref}[1]{Eq.~\ref{#1}}
\newcommand{\secref}[1]{Sec.~\ref{#1}}
\newcommand{\figref}[1]{Fig.~\ref{#1}}
\newcommand{\tabref}[1]{Tab.~\ref{#1}}
\newcommand{\corrfunc}{C}

\newcommand{\convolve}{\circledast}
\newcommand{\modulation}{i}
\newcommand{\demodulation}{s}
\newcommand{\gaussian}[1]{\mathcal{G}(#1)}

\newcommand{\depth}{\Gamma}
\newcommand{\phase}{\varphi}

\newcommand{\doidepth}{\depth_0}
\newcommand{\doiphase}{\phase_0}
\newcommand{\doiphaseshift}{\theta_G}

\newcommand{\phasedepth}{\phase_\depth}
\newcommand{\sensitiveregion}{\Delta\Gamma}
\newcommand{\sensitiveregionphase}{\Delta\phasedepth}

\newcommand{\rawfraction}{\Psi}
\newcommand{\rawfractionlong}{\frac{\corrfunc_0(\pixel) - \corrfunc_2(\pixel)}{\corrfunc_1(\pixel) - \corrfunc_3(\pixel)}}

\newcommand{\fwhm}{\text{FWHM}}

\begin{document}

\title{A new operation mode for depth-focused high-sensitivity ToF range finding}

\author{Sebastian Werner$^{1}$, Henrik Sch\"afer$^{2}$ and Matthias Hullin$^{1}$\\
\\
{\small $^{1}$University of Bonn, Department of Computer Graphics}\\
{\small $^{2}$Sony Europe B.V., European Technology Center}
}
\date{}
\maketitle

\begin{abstract}
We introduce pulsed correlation time-of-flight (PC-ToF) sensing, a new operation mode for correlation time-of-flight range sensors that combines a sub-nanosecond laser pulse source with
a rectangular demodulation at the sensor side. In contrast to previous work, our proposed measurement scheme attempts not to optimize depth accuracy over the full measurement:
With PC-ToF we trade the global sensitivity of a standard C-ToF setup for measurements with strongly localized high sensitivity -- we greatly enhance the depth resolution for the acquisition of scene features around a desired depth of interest.
Using real-world experiments, we show that our technique is capable of achieving depth resolutions down to 2\,mm using a modulation frequency as low as 10\,MHz and an optical power as low as 1\,mW. 
This makes PC-ToF especially viable for low-power applications. 
\end{abstract}

\section{Introduction}\label{sec:introduction}
\textit{Time-of-flight} (ToF) range finding setups support a vast amount of applications, ranging from robotics closely tied with exploration and automated manufacturing to motion capture and 3D mapping, as well as biometrics~\cite{Kolb2009}. 
They are all connected by the common need for truthful representations of the three-dimensional environment. As a consequence, all applications share the desire for both -- high spatial resolution as well as precise depth estimation. 
Thanks to advances in sensor technology, the former rises with every generation of sensors, whereas the depth resoltuion depends on deisgn choices, such as the time and power budget of any such sensor and hence fundamentally limited by noise. Especially for low \textit{signal-to-noise ratio} (SNR) measurements, accurate detection of distances becomes a challenge that received a lot of attention from the scientific community. 
There exists a number of range finding approaches based on ToF measurements, which can be divided into two classes that differ in both hardware requirements and reconstruction techniques.
\paragraph{Direct pulsed time-of-flight range finding.}
Direct pulse-based ToF systems (P-ToF)~\cite{Goldstein1967,Koechner1968} were the first ToF systems to be employed for range finding purposes. These setups emit a single short (pico-/nanosecond) laser light pulse into the scene. The sensor then receives a delayed pulse after a certain travel time. The time delay between emission and acquisition is directly proportional to the distance travelled and subsequently the depth of the scene. Recent pulse-based systems rely on the determination of the pulse-shape, altered by scene traversal~\cite{Iddan2001,Yahav2007} and implementation of a fast image shutter in front of the sensor chip. 
Use cases are as diverse as acquiring images~\cite{Kirmani2009} or object motion~\cite{Pandharkar2011} in an ``around the corner'' setting, measuring 3D shape~\cite{Velten2012} or separation of light transport components~\cite{Wu2012}. 
The simplicity of the underlying concept comes at the cost of elevated hardware requirements, enabling the measurement of time delays in the order of picoseconds in low SNR scenarios. 
Due to these limitations, such systems often consist of a single-pixel sensor only and require time-consuming pixel-wise scanning of the scene. Despite those shortcomings, 
the strength of P-ToF systems lies in their high depth resolution. 
\paragraph{Correlation time-of-flight range finding.}
To alleviate the need for fast and costly hardware, \textit{amplitude-modulated continuous-wave} (AMCW) ToF systems have been developed that consist of temporally modulated light sources and sensors \cite{Schwarte1997,Lange2000}. At the core 
of these \textit{correlation time-of-flight} (C-ToF) setups lie the modulation (at the light source) and demodulation (at the sensor) functions, that are used to code and decode the illumination signal. 
Current C-ToF setups utilize sinusoidal or square coding functions. Upon scene traversal, the amplitude modulated illumination undergoes a phase shift with respect to the original signal emitted by the light source. This phase shift 
is proportional to the traveled distance and is acquired using a correlation measurement between the emitted and received signal. As modulation and 
demodulation function are periodic, these measurements implicitly are limited to the so-called \textit{unambiguity range}, which depends on the frequency of the modulation signal. Our approach relies on a homodyne setup, where the frequency of the modulation and demodulation signals are equal.
\paragraph{Depth resolution enhancements for C-ToF systems.}
In comparison to P-ToF systems, correlation-based ToF systems exhibit considerably lower depth resolution. This is due to the fact that C-ToF systems rely on single- or few-frequency signals such as sinusoidal or triangular~\cite{Ferriere2008} modulation and demodulation signals, which in turn renders the depth estimation more prone to errors from measurement noise~\cite{Buttgen2008}. In recent years a great amount of research has been done to mitigate effects from higher harmonics of 
such modulation-demodulation signal pairs~\cite{Payne2008,Payne2010b}. Dual-frequency setups try to enhance depth resolution without the loss of unambiguous measurement range by combining high- and low-frequency measurements~\cite{Jongenelen2010,Jongenelen2011}. More recently, Gupta et al.~\cite{Gupta2018} presented a framework for general C-ToF range finding, which allows for the simulation and computation of the depth resolution performance for arbitrary modulation-demodulation signal pairs. In addition, they also used their framework to develop an optimized Hamiltonian coding function, which achieves depth resolutions below 1\,cm over the full ambiguity range. 
Closely tied to the work presented in this paper, Payne et al.~\cite{Payne2011} discuss the optimal choice of the duty cycle for the chosen illumination signal for the special case of sinusoidal and square illumination modulation. They point out 
that a reduction of duty cycle results in an increased peak power and thus better SNR for the illumination signal, whereat the linear relation used for phase estimation is violated by a change of frequency content of the signal. 
All these approaches share the desire for improvements of the depth sensitivity over the full unambiguity range, which are inherently deemed to result in a tradeoff due to their respective relations to the modulation frequency. This is due to 
the fact, that the limited bandwidth of the illumination signal(s) directly relates to the depth variations the procedure can truthfully distinguish. 
With \emph{pulsed correlation time-of-flight sensing} (PC-ToF), we propose a dual-measurement scheme that explicitly makes use of this information content. 
This novel hybrid approach combines the high depth resolution and noise resilience of P-ToF with the low-cost hardware of C-ToF setups at the cost of ambiguity: We replace the (continuous) modulation function with pulsed illumination but maintain a continuous demodulation signal. This way, we make use of higher harmonics in the modulation signals, previously treated as artefacts. 
The two steps of PC-ToF are:
\begin{enumerate}
 \item Obtain a rough depth estimate with standard C-ToF range finding methods and select a \textit{depth of interest} (DOI) for close inspection. 
 \item precisely measure the depth around a user-specified DOI with depth resolutions down to 2\,mm, utilizing our hybrid approach.
\end{enumerate}
The power consumption of C-ToF systems is dominated by optical output power and sensor modulation. Its light efficiency and the use of slow modulation frequencies make PC-ToF especially suitable for low power applications, for example in mobile hardware.
%
%
\begin{figure}[t]
 \includegraphics[width=\linewidth]{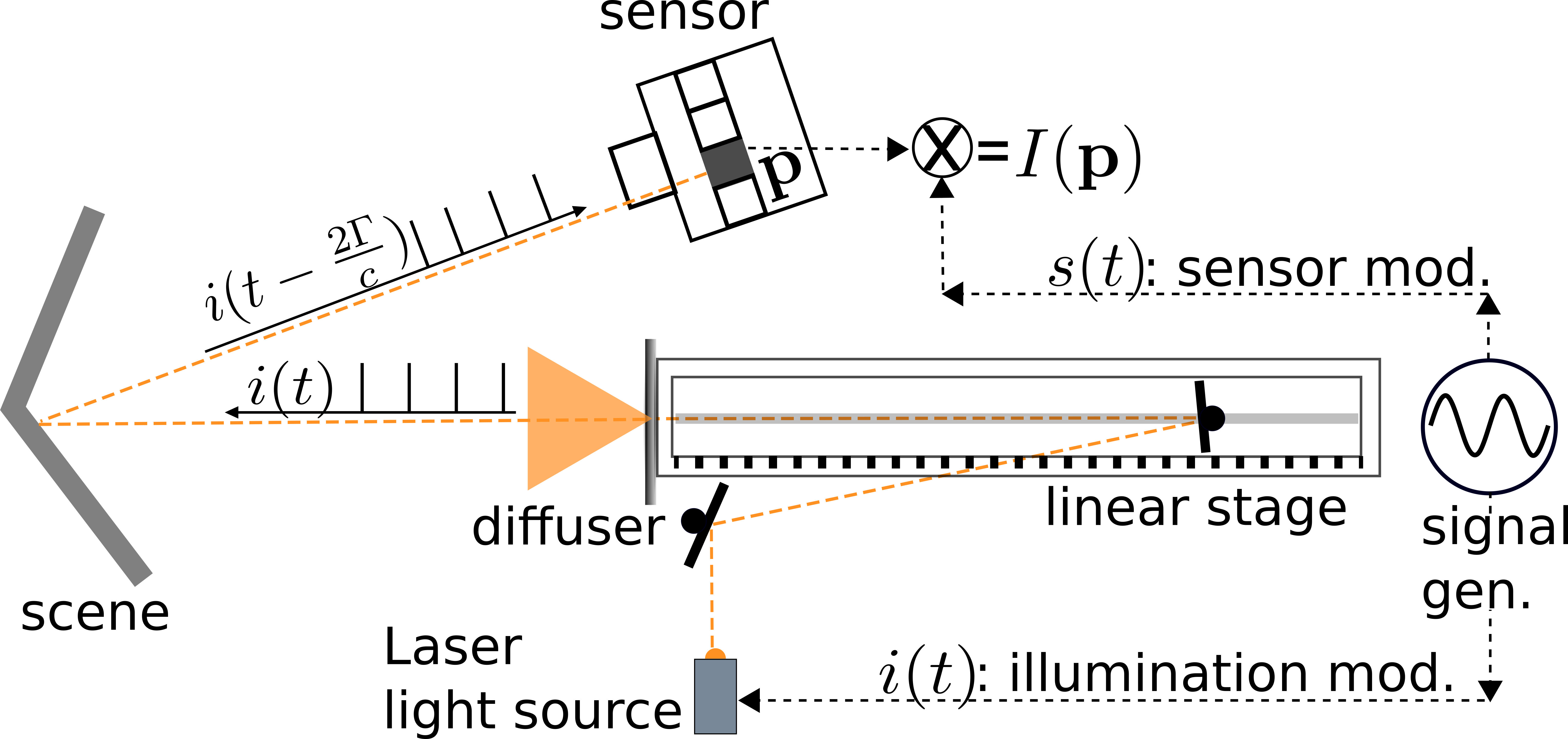}\\
 \includegraphics[width=\linewidth]{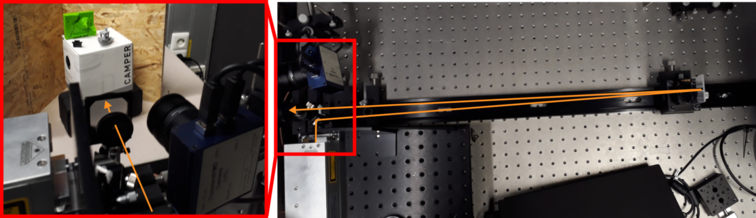}
 \caption{\textbf{Upper panel}: Schematic visualization of our correlation ToF setup. A signal generator drives both -- modulation of the illumination and the sensor gain. In pulsed operation mode, the laser light source emits a pulse chain  $i(t)$. Otherwise, we utilize a sinusoidal illumination modulation. The light is guided onto a mirror on a linear stage, which allows to control the distance traveled, required for our validation procedure. For 
 uniform illumination of the scene, the light is focused onto a diffuser.Per pixel $\pixel$, the sensor then retrieves a shifted version of the illumination signal, which is correlated with the sensor modulation $s(t)$.
 \textbf{Lower panel}: Pictures of our lab setup, as indicated in the upper panel.}
 \label{fig:CToF-scheme}
 \end{figure}
\section{Correlation time-of-flight image formation}\label{sec:theory}
We will briefly revisit the image formation model for correlation ToF as described 
in~\cite{Gupta2015,Gupta2018} and, for simplicity, adapt their notation: 
Correlation time-of-flight setups (see~\figref{fig:CToF-scheme}) consist of an 
amplitude-modulated light source and a gain-modulated sensor. We start by defining 
the modulation functions of the light source $\modulation(t)$ and sensor 
gain $\demodulation(t)$ respectively. Like~\cite{Gupta2018,Freedman2014,Heide2013,Kadambi2013}, 
we assume the absence of any indirect or multi-bounce light, which allows us to 
describe the scene response as a single scattering event at the precise depth $\depth$. 
This results in a shift of the modulation function $\modulation(t)$. In general, the 
\textit{irradiance} $E(\pixel, t)$ that arrives at pixel $\pixel$ can then be 
described as 
\begin{equation}
\label{eq:pixel-radiance}
E(\pixel, t) = E_a(\pixel) + E_c(\pixel) \, \modulation(t - 2\depth/c)
\end{equation}
where $c$ is the speed of light, $E_a(\pixel)$ denotes the ambient light component and $E_c(\pixel)$ is the 
mean pixel irradiance due to the modulated light, encoding the optical properties of the scene.
The shifted normalized illumination modulation is described by $\modulation(t - 2\depth/c)$. The sensor then records the \textit{pixel intensity}
\begin{align}
\label{eq:pixel-brightness}
 I(\pixel) &= \int_{0}^{\tau} E(\pixel, t) \demodulation(t)\mathrm{d}t\nonumber\\
	   &= I_a(\pixel)+E_c(\pixel)\int_{0}^{\tau} \modulation \left( t- 2\depth(\pixel)/c \right) s(t)\,\mathrm{d}t\mathrm{,}
\end{align}
where $\tau$ is the exposure time and 
$I_a(\pixel)=\int_{0}^{\tau}E_{a}(\pixel)\demodulation(t)\mathrm{d}t$ is the incident ambient light.
Eq.~\ref{eq:pixel-brightness} can be understood as a cross-correlation function. 
From this, we define the \textit{normalized correlation function}
\begin{equation}
 \label{eq:normalized-correlation-function}
\corrfunc(\depth) = \int_{0}^{\tau} \modulation \left( t-2\depth/c \right)\demodulation(t)\,\mathrm{d}t \mathrm{.}
\end{equation}
This way, we are able to express the full image formation process of correlation time-of-flight imaging via the \textit{image formation equation}
\begin{equation}
  \label{eq:image-formation}
 I(\pixel) = E_c(\pixel) \corrfunc(\depth)+I_a(\pixel).
\end{equation}
This equation reveals three unknowns $E_c(\pixel), \depth, I_a(\pixel)$ which have to be determined pixel-wise. We require $K\ge3$ measurements 
or samples of the correlation function $\corrfunc_i(\depth)$ for $i \in \{0,\ldots,K\}$. These measurements are commonly realized by inserting an additional phase shift $\theta_i$ into the demodulation function, such that
\begin{equation}
  \label{eq:demodulation_signal_2}
  \demodulation(t) \rightarrow \demodulation_i\left(t + \theta_i/\omega \right);\,\theta_i \in [0,2\pi)
\end{equation}
and data acquisition is performed for $K$ equally spaced phases. 
Gupta et al.~\cite{Gupta2018} continue to develop a \textit{depth precision measure} $\bar{\chi}_{c}$, which encodes the average depth accuracy depending on the average optical properties encoded in $E_{c,\mathrm{mean}}$  as well as the 
noise standard deviation $\Omega = \sqrt{\sum_{i=1}^K\sigma_i^2}$ (assumed to be constant) for $K$-tap correlation ToF measurements as
\begin{equation}
  \label{eq:depth-precision}
 \bar{\chi}=\frac{ E_{c} }{ \Omega\depth_\mathrm{range} } \int_{\depth}{\sqrt{\sum_i \left( \partial \corrfunc_{i}(\depth)/\partial \depth \right)^{2}}d\depth} \mathrm{,}
\end{equation}
where $\depth_\mathrm{range}$ is the unambiguous depth range.
%
 %
\begin{figure*}
\begin{minipage}{0.33\linewidth}
  \includegraphics[width=\linewidth]{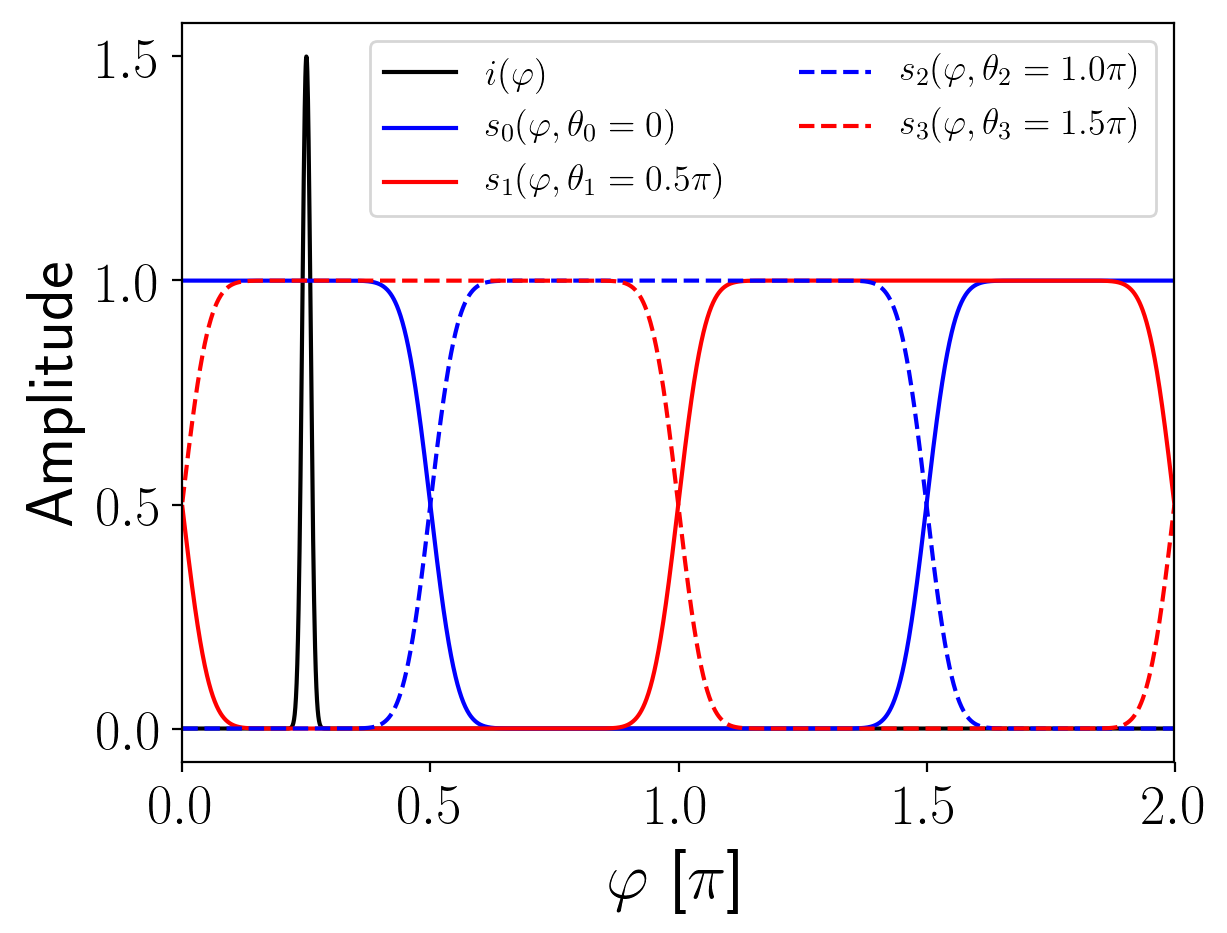}
\end{minipage}
\begin{minipage}{0.33\linewidth}
  \includegraphics[width=\linewidth]{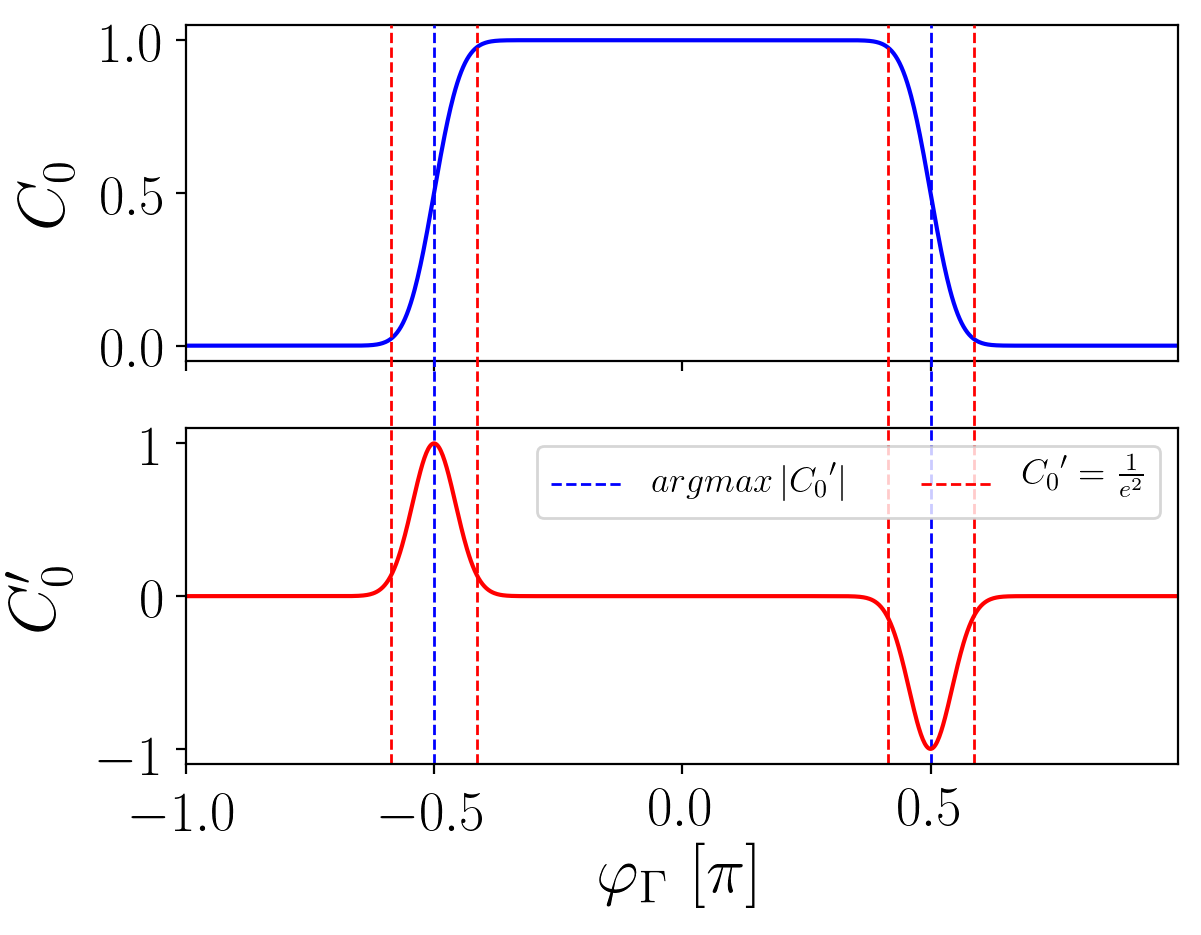}
\end{minipage}
\begin{minipage}{0.33\linewidth}
 \includegraphics[width=\linewidth]{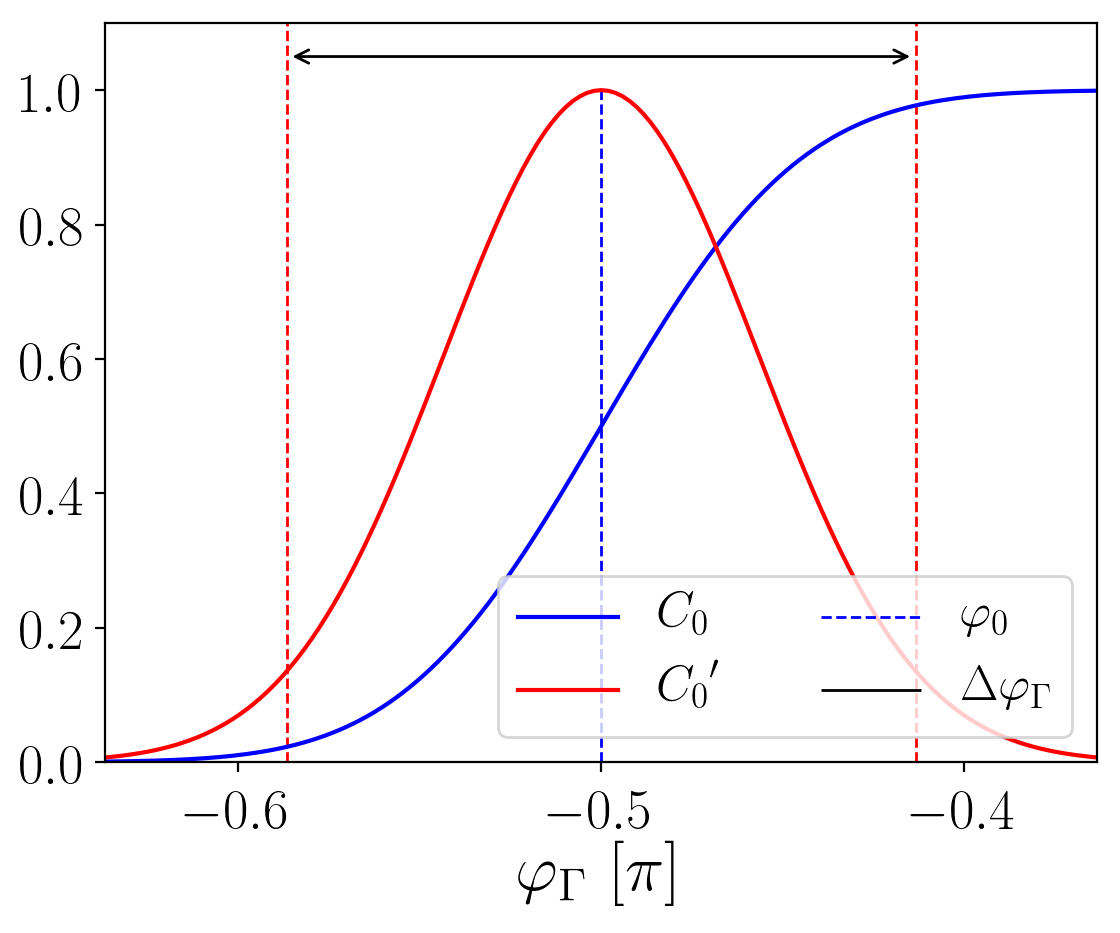}
\end{minipage}
  \caption{
  \textbf{Left}: Signals of the pulsed ToF approach. Note that the modulation signal $\modulation(\phase)$ is rescaled for visualization. The demodulation signals are plotted for the denoted shifts applied, 
  corresponding to a standard 4-tap ToF measurement.
\textbf{Middle}: Correlation function $\corrfunc_0(\phasedepth)$ and its first derivative $\partial \corrfunc_0/\partial \phasedepth$ in dependence on the phase (and hence, depth). Given a chosen phase shift $\theta_i$, only certain depths lie within the \textit{sensitive range}, here indicated as the region between two dashed red vertical lines surrounding an extremum. Only depths corresponding to phases within this range lead to reliable depth measurements. 
  The \textit{phase of maximum sensitivity} is reached when $\partial \corrfunc_0/\partial \depth$ reaches an extremum, denoted by the dashed blue line. We denote the corresponding depth as the \textit{depth of interest} $\doidepth$, onto which we are able to focus by shifting $\theta_i$. \textbf{Right}: Close-up of $\corrfunc_0(\phasedepth)$ and $\partial \corrfunc_0/\partial \phasedepth$ around the first extremum $\doiphase$. To restrict the sensitive range to non-negligible values, we define it as the beam width of the Gaussian.
  }
  \label{fig:PToF-signals}
\end{figure*}
\section{A new operation mode for time-of-flight range finding}
Our foremost aim is to increase the depth sensitivity not on a global scale (over the full ambiguity range) but locally. This allows us to select a certain \textit{depth of interest} (DOI), around which we can retrieve the depth information of 
the scene with high accuracy. 
From \eqref{eq:depth-precision} we directly see that depth sensitivity depends on the gradient of the (normalized) correlation signal. Ideally, $\partial \corrfunc/\partial \depth \to \infty$ which would result in a vanishing rise time $T_\text{Rise} = t(\max(\corrfunc)) - t(\min(\corrfunc))\to 0$. We base our considerations on the idealized case of a combination of pulse trains (Dirac comb) for our modulation signal and using a rectangular demodulation signal 
on the sensor side, both with frequency $\nu$. To unify considerations and clarify, that we are limited to exactly one period of the modulation and demodulation signals before ambiguities arise, we switch the integration variable to phase $\phase$ via 
\begin{equation}
 \phase = \omega t;\, \omega=2\pi\nu;\, \phasedepth = \frac{2\omega\depth}{c} \mathrm{.}
\end{equation}
\\
We describe our (real) modulation and demodulation signals as a chain of Gaussian pulses and a smoothed rectangular signal chain respectively (cf.~\figref{fig:PToF-signals}). 
The modulation signal is then described as the convolution $\convolve$ of a Dirac comb with a Gaussian $\mathcal{G}$ with standard deviation $\sigma_{M}$,
\begin{equation}
 \label{eq:modulation_signal_1}
 \modulation(\phase) = \sum_{n} \left[ \delta(\phase - n\,2\pi - \phasedepth) \convolve \gaussian{\phase,\sigma_M} \right] \mathrm{,}
 \end{equation}
where the \textit{pulse width} is assumed to equal the FWHM. 
We model the demodulation signal $\demodulation$ as a square signal onto which we apply a Gaussian smoothing kernel to account for non-vanishing rise times:
\begin{equation}
 \label{eq:demodulation_signal_1}
 \demodulation_i(\phase) = \sum_{n} \left[ \rect \left( \frac{\phase - n\,2\pi + \theta_i}{\pi} \right) \convolve \gaussian{\phase,\sigma_D}\right]\mathrm{.}
\end{equation}
To compute the correlation function (\eqref{eq:normalized-correlation-function}), we utilize the fact that the convolution of two Gaussians yields another Gaussian function with
$\sigma=\sqrt{\sigma_D^2 + \sigma_M^2}$. Assuming the pulse width being smaller than the period of the modulation ($\fwhm << \omega T$), we obtain
\begin{align}
 \label{eq:correlation_function}
  \corrfunc_i(\phasedepth) =& \frac{1}{\omega} \int_{0}^{2\pi} \rect \left( \frac{\varphi + \theta_i - \phasedepth}{\pi} \right) \convolve \gaussian{\phase,\sigma}\,d\varphi \nonumber\\
		      =& \frac{1}{\omega}\frac{\sqrt{\pi}}{2\sqrt{a}}\{\erf(\sqrt{a}\varphi_{2})-\erf(\sqrt{a}\varphi_{1})\nonumber\\
	   \varphi_{2,1} =&  \pm\frac{\pi}{2} + \phasedepth - \theta_i;\,a = \frac{1}{2\sigma^2}\mathrm{.}
\end{align}
\paragraph{Depth sensitivity.}
The depth sensitivity (\eqref{eq:depth-precision}) is driven by the gradient of the normalized correlation function
\begin{align}
\frac{\partial \corrfunc_i(\phasedepth)}{\partial \phasedepth} =& \frac{1}{\omega} \left\{ \exp(-a (\phasedepth - \theta_i + \frac{\pi}{2})^2) \right. \nonumber\\
    & \left. \;\; -\exp(-a (\phasedepth - \theta_i - \frac{\pi}{2})^2)\right\}\mathrm{,}
\end{align}
which essentially are two Gaussians located at 
\begin{equation}
\label{eq:maximum-sensitivity-phase}
	\phasedepth=\theta_i \pm\frac{\pi}{2}\mathrm{,}
\end{equation}
one with negative, the other with positive amplitude (see \figref{fig:PToF-signals}, middle). 
These Gaussians indicate that the maximum (absolute) gradient of the correlation function $\corrfunc(\depth)$ is achieved at this local maximum and minimum respectively, which depend on the value of $\phasedepth$ and hence the distance 
towards an observed object. This means that our PC-ToF approach exhibits strong sensitivity in a narrow range around a specific phase, the \textit{phase of maximum sensitivity} $\doiphase$. 
\paragraph{The depth of interest (DOI).}
\begin{figure*}
\begin{minipage}{0.33\linewidth}
 \includegraphics[width=\linewidth]{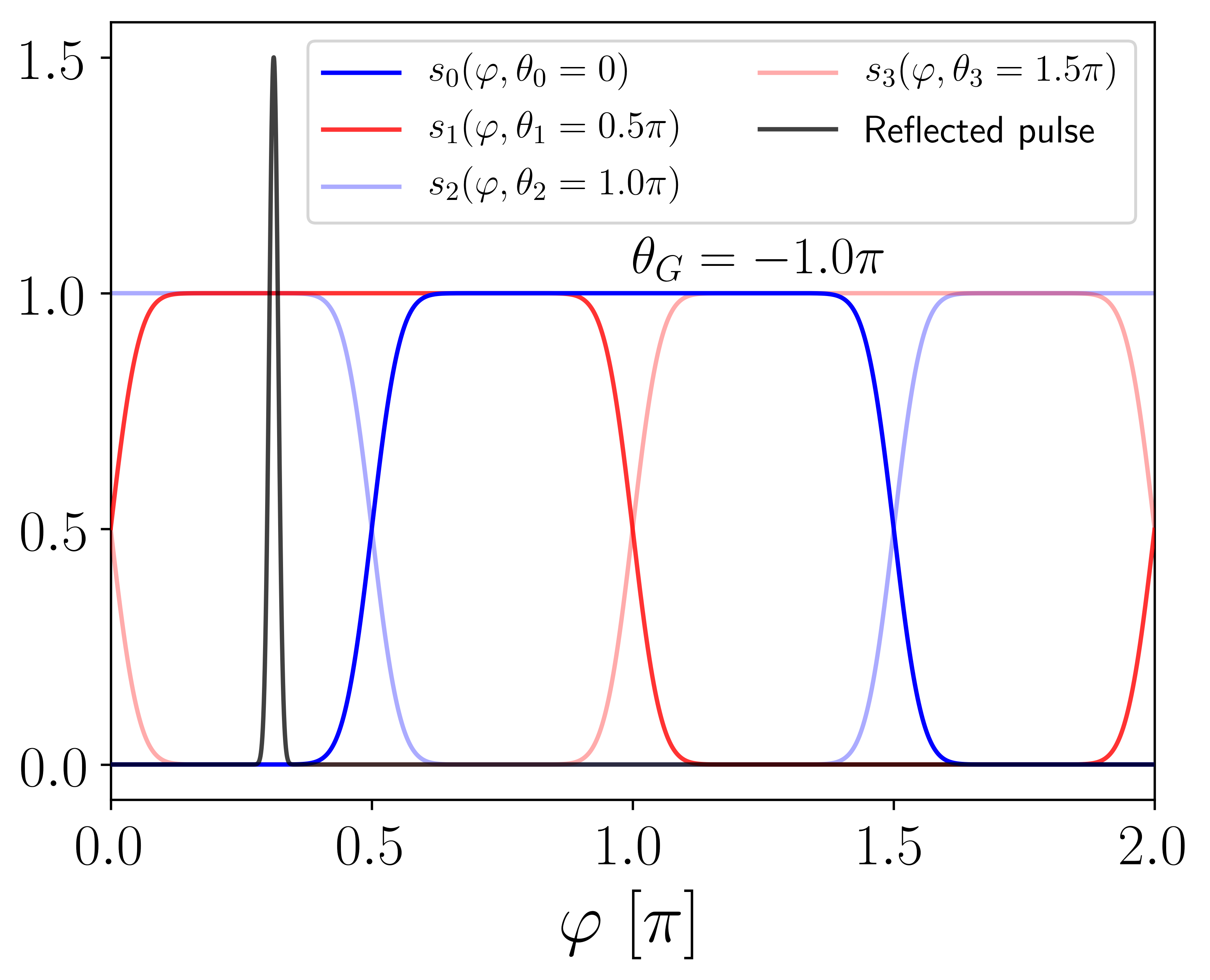}
\end{minipage}
\begin{minipage}{0.33\linewidth}
 \includegraphics[width=\linewidth]{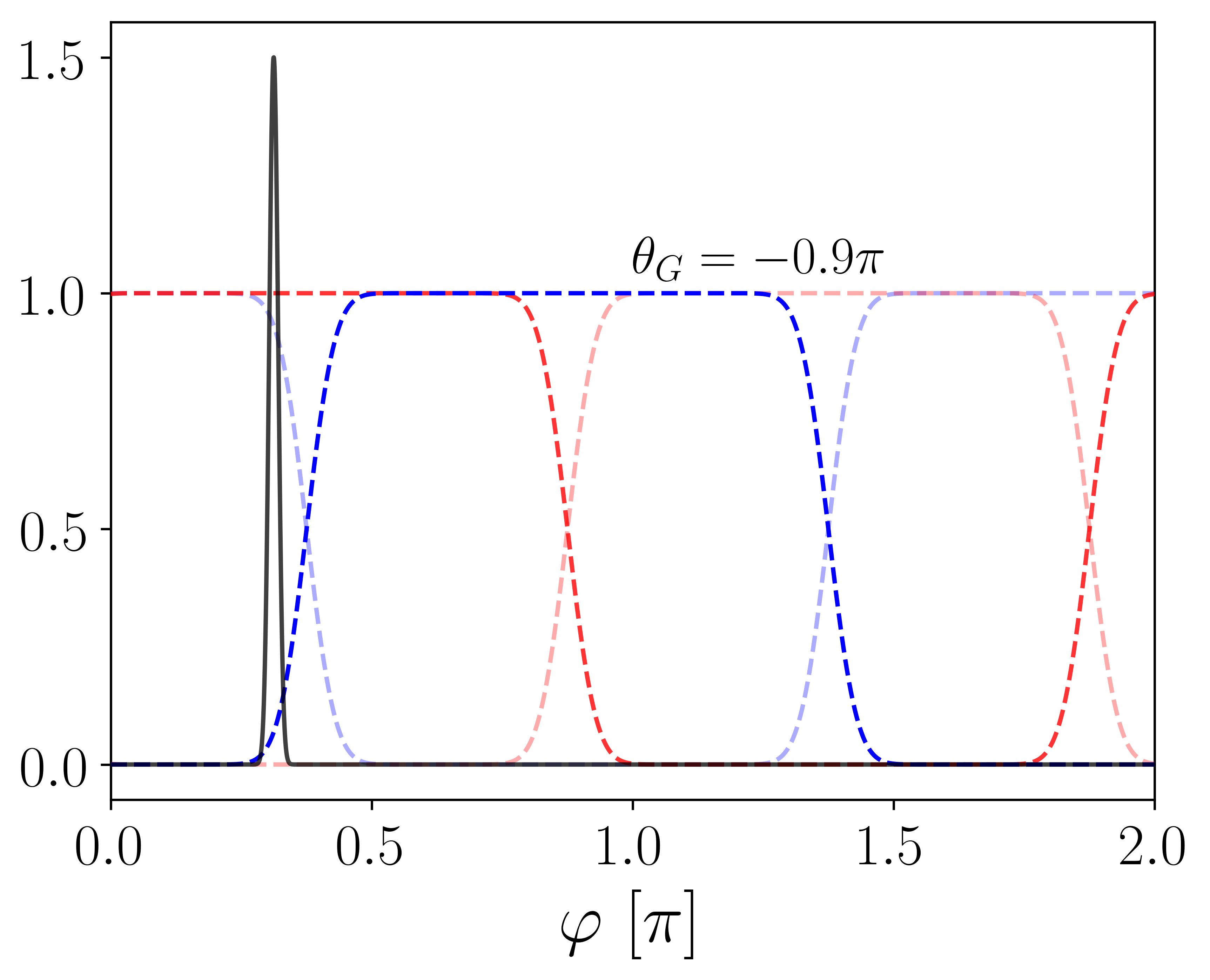}
\end{minipage}
\begin{minipage}{0.33\linewidth}
 \includegraphics[width=\linewidth]{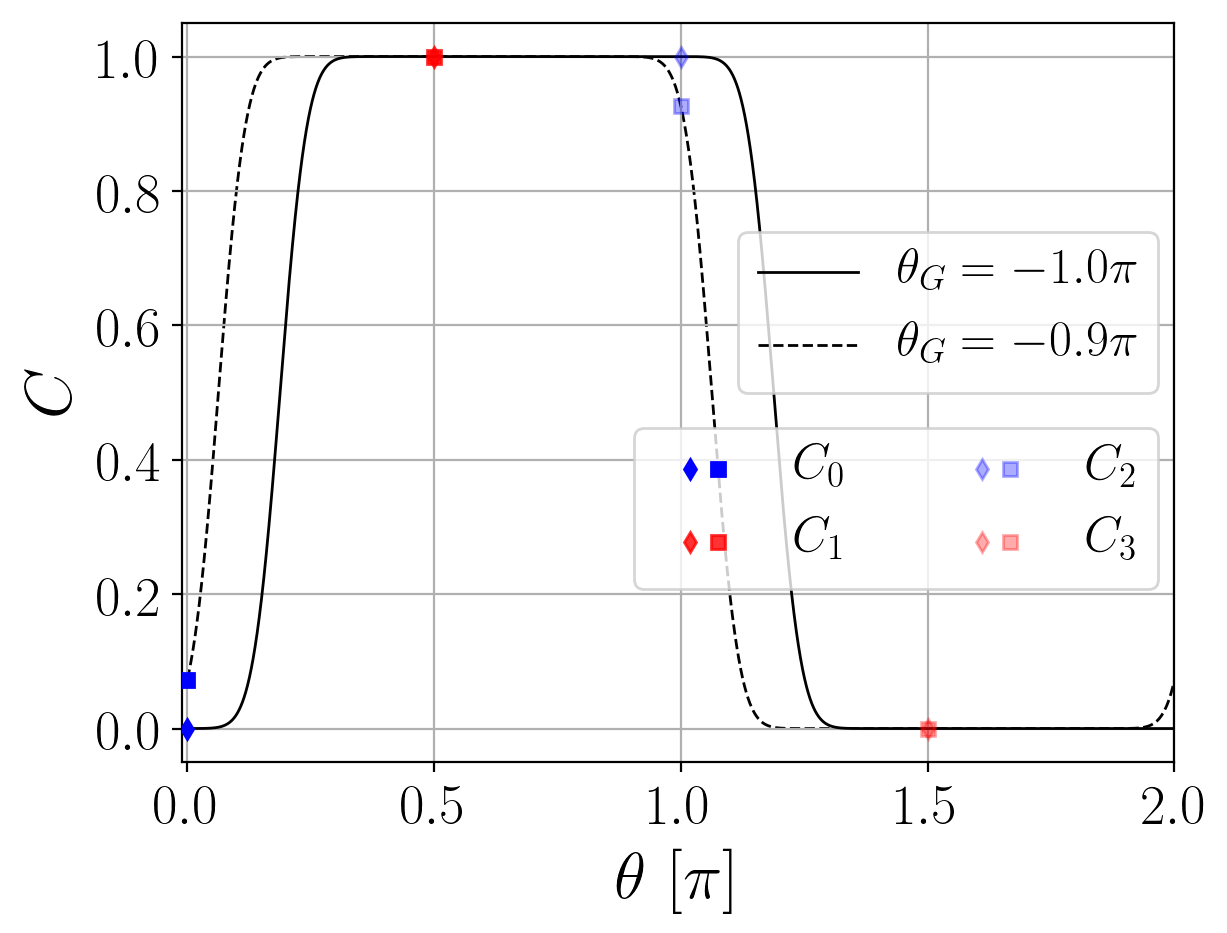}
\end{minipage}
\caption{Demodulation signals and reflected pulse and corresponding correlation function measured in a single pixel. The demodulation signals are color coded to represent their respective phase shift for a 4-tap sampling procedure.
	 \textbf{Left}: The reflected pulse coincides with the lower and upper plateaus of $\demodulation(\phase)$ due to a bad choice of $\doidepth$ and hence $\doiphaseshift$. The 4 measured samples are obtained at maximum and minimum value respectively (see \textbf{right} panel, solid line). 
	 This way, ambiguity arises and no certain phase can be reconstructed, as nearby depths (or reflected pulses) will yield the same result. 
	 \textbf{Middle}: We choose a rough estimate for the DOI $\doidepth$ and adjust the phase shift $\doiphase$ by applying $\doiphaseshift$ such that the reflected pulse lies within the \textit{sensitive range}. This results in a coincidence of reflected pulse and rising signal edge of $\demodulation_0(\phase)$. The 4 measured samples now refer to a unique phase (see \textbf{right} panel, dashed line) and a small change in measurement value will result in a large change of the phase estimate. 
	 }
\label{fig:correlation-sampling-combined}
\end{figure*}
From \figref{fig:PToF-signals} (middle, right) it becomes clear that only measurements with a specific depth ($\doidepth(\doiphase)$) can be made at maximum sensitivity and produces meaningful results. 
Relation \eqref{eq:maximum-sensitivity-phase} reveals, that a sensible choice of $\theta_i$ allows to shift the correlation function such, that the extrema are located at the desired phase $\phasedepth = \doiphase$ and hence depth $\doidepth$. 
This depth we call the \textit{depth of interest} (DOI). In a $K$-tap measurement system it is not directly clear which of the phase shifts $\theta_i$ should be chosen such that $\partial \corrfunc/\partial \phasedepth$ exhibits an extremum. 
For simplicity we will choose $\theta_0$ which introduces a global phase shift $\doiphaseshift$ as the $\theta_i$ are equally spaced: 
\begin{equation}
\doiphase = \phasedepth \mp \frac{\pi}{2} - \theta_0 = \frac{2\omega\doidepth}{c} - \frac{\pi}{2};\, \theta_i \rightarrow \theta_i + \doiphaseshift
 \label{eq:doi}
\end{equation}
\paragraph{Sensitive range.}\label{sec:sensitive-range}
The phase of maximum sensitivity is restricted to one particular value and corresponding depth, whereat measurements within a surrounding phase interval also benefit from increased depth sensitivity. 
We call this interval the \textit{sensitive range} $\sensitiveregion$ and corresponding phase $\sensitiveregionphase$. Mathematically it can be described as the interval with nonzero first derivative, i.e., $\partial\corrfunc_0/\partial\phasedepth \neq 0$. 
We estimate this interval as the rise time of the signal edge surrounding $\doiphase$. In general we would focus on the rising signal edge (see \figref{fig:PToF-signals}, right) and ask for the roots of the first derivative. 
As an example consider a sinusoidal correlation function, where the phases corresponding to the sensitive range are exactly $\pi$ apart, resulting in full sensitivity over the ambiguity range. On the other hand, the derivative of our correlation function consists of two Gaussians with nonzero value everywhere. We therefore define the sensitive range as the interval bounded by the points where 
the Gaussian reaches $1/e^2$ of its peak, often also called the \textit{beam width}.
\begin{equation}
\label{eq:sensitive-range}
 \sensitiveregionphase = 4\sigma;\, \sensitiveregion = \frac{\sensitiveregionphase\,c}{2\omega}
\end{equation}
For illustration, consider a PC-ToF measurement which has a true depth of $\depth$. To achieve maximum sensitivity, the ideal solution would be to set the DOI to the exact depth and acquire the necessary phase shift $\doiphaseshift$. 
Physically this is the case when the reflected illumination pulse coincides with the rising edge of the demodulation signal. 
In contrast, the reflected pulse coincides with one of the plateaus of the demodulation signals (\figref{fig:correlation-sampling-combined} left) if the true depth $\Gamma$ and $\Gamma_0$ differ too strongly. 
The measured $\corrfunc_i$ do not change upon small changes of the depth -- the measured values are outside the sensitive range $\sensitiveregion$. 

However, a priori the exact depth value is unknown. To still achieve a high resolution depth measurement we will choose the DOI $\Gamma_0$ such that it is close to the (unknown exact) depth $\Gamma$. 
Having an estimate for $\Gamma_0$ that lies within the sensitive range results in a coincidence of reflected pulse and signal edge (\figref{fig:correlation-sampling-combined} right) -- 
the measurement operates at increased, albeit not necessarily maximum sensitivity and yields an accurate result for $\Gamma$. 
\subsection{Hardware}
Our approach describes an additional mode of operation for existing correlation ToF range finding setups, which requires certain hardware characteristics to be available.\\
\textbf{First}, we require a correlation ToF sensor. These devices are either externally modulated by a high-frequency signal or employ their own signal generator for this purpose. We utilize a PMD CamBoard nano (based on their 19k-S3 sensor) with external DDS modulation source at 10\,MHz \cite{Heide2013}, which also triggers the laser source. 
\textbf{Second}, we require the light source to emit pulses with the given modulation frequency and narrow pulsewidth. To this end we utilize an Omicron QuixX laser with pulse width $\fwhm \le 500$\,ps.\\
\textbf{Third}, for calibration we require a phase shift to be applied to either the modulation or demodulation signal. This phase shift needs to be adjustable with as high an accuracy as possible, as this affects the final resolution of the range imaging system. The modulation source allows setting the phase with 14 bits precision, leading to phase steps as small as $\Delta\phase = 2\pi/2^{14}$.
The parameters for our measurements, as well as the specific hardware used, can be found in \tabref{tab:hardware}.
\begin{table}[]
\centering
\caption{Parameters of the hardware used for our measurement setup (cf.~\figref{fig:CToF-scheme})}
\resizebox{\linewidth}{!}{
\begin{tabular}{|l|l|l|l|l|l|}
\hline
\multicolumn{2}{|l|}{\textbf{Laser light source}} & \multicolumn{2}{l|}{\textbf{Sensor}}        & \multicolumn{2}{l|}{\textbf{Lens}}             \\ 
\multicolumn{2}{|l|}{Omicron QuixX 852-150} & \multicolumn{2}{l|}{PMD 19k-S3} & \multicolumn{2}{l|}{Fujinon HF35SA-1} \\ \hline
Wavelength            & 852 nm              & Resolution          & 160x120      & Focal length           & 35mm         \\ \hline
Pulse width (FWHM)    & \textless{}500 ps   & Frequency           & 10 MHz       & Aperture               & f/2.0        \\ \hline
Average power         & \textless 1mW       & Shutter time        & 1 ms         & \#Acquisitions         & 25           \\ \hline
Sensitive range $\sensitiveregion$ & $\lessapprox0.75$\,m &	  & 		 &			  &		 \\ \hline
\end{tabular}
}
\label{tab:hardware}
\end{table}
\subsection{Setup and measurement procedure}\label{sec:setup}
\figref{fig:CToF-scheme} shows a schematic illustration as well as pictures of our setup: The pulsed laser illumination is guided onto a mirror mounted on a linear stage before being reflected back 
onto a diffuser for uniform illumination of the scene. The linear stage allows to control the distance travelled which directly translates to a proportional phase shift. This is equivalent to adding a phase shift in hardware and 
is used for validation only, but could in principle also be utilized for calibration. The sensor observing the scene then retrieves a delayed version of the illumination signal, which is correlated with the demodulation signal 
on a per-pixel level. 
\paragraph{Depth reconstruction.}
The CamBoard nano, as most available C-ToF systems, employs a four-tap measurement procedure that acquires four samples $\{\corrfunc_0(\pixel), \corrfunc_1(\pixel), \corrfunc_2(\pixel),\corrfunc_3(\pixel)\}$ 
of the correlation function per pixel $\pixel$ measured using demodulation functions shifted by $\theta_i \in \{0,\frac{1}{2}\pi, \pi, \frac{3}{2}\pi\}$. Instead of disclosing these four values, the CamBoard nano returns 
the differences of samples separated by $\Delta\phase = \pi$. For C-ToF systems utilizing sinusoidal modulation and demodulation signals it can be shown~\cite{Schwarte1997} that the unknown phase $\phasedepth(\pixel)$ corresponding to 
the range $\depth(\pixel)$ can be computed as
\begin{align}
 \label{eq:atan-reconstruction}
 \phasedepth  &= \frac{2\omega\depth}{c} = \atan\left(\rawfraction\right)\nonumber\\
 \rawfraction &= \rawfractionlong
\end{align}
and we denote the argument $\rawfraction$ as the \textit{raw fraction}. This expression has two major benefits: First, the result of the differences is independent on ambient light $I_a(\pixel)$ and second, the fraction of the two differences cancels out the scene dependent factor $E_c(\pixel)$. Still, \eqref{eq:atan-reconstruction} is only valid for sinusoidal signals and results in strong systematic errors~\cite{Payne2008} for non-harmonic correlation functions such as ours. 
Instead, we will rely only on measurements of the raw fraction in dependence of a chosen depth of interest $\doidepth$ and phase shift $\doiphaseshift$.  
\paragraph{Calibration.}
\begin{figure*}
  \includegraphics[width=\linewidth]{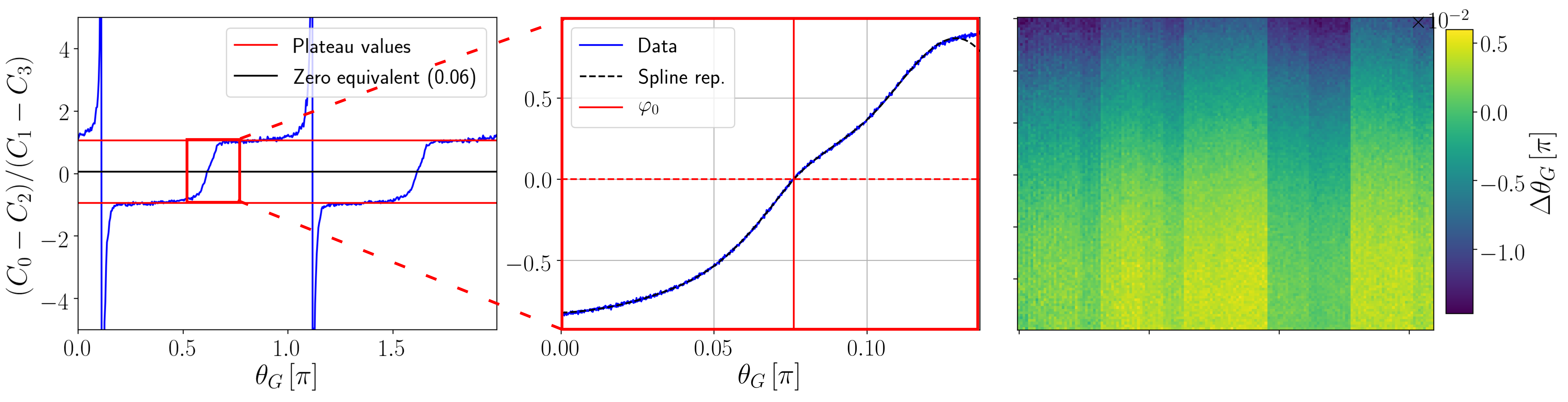}
 \caption{
 \textbf{Left}: Measured raw fraction $\rawfraction$ for 512 equally spaced phase shifts $\doiphase \in [0, 2\pi)$. We estimate the values for the upper and lower plateaus of the correlation function (red horizontal lines). 
 From the plateau values we compute a zero-crossing equivalent value, which is exactly half the difference between the plateaus. We further estimate the sensitive range, here denoted by the red rectangle. 
 \textbf{Middle}: We perform an additional measuremetn of $\rawfraction$ for phases $\doiphase$ within the sensitive range enclosing the rising signal edge, which is performed at the highest possible accuracy in terms of $\doiphase$ (14 bit).  
 The data acquired exhibits noise, which leads to ambiguities when used for a lookup table. Instead, we fit a continuous spline representation to circumvent that issue. This way, we obtain a $\doiphase$-$\rawfraction$ mapping that allows to 
 estimate the offset a measurement exhibits from a reference phase.
  \textbf{Right}: As we can only chose a single depth of interest via $\doiphase$ per measurement, we obtain the phase of maximum sensitivity for each single pixel and subtract it from the median over all pixels.  
  This way, we obtain a calibration mask that employs a per-pixel phase correction with respect to $\doiphase$.
 }
 \label{fig:measurement_calibration}
\end{figure*}
In contrast to the simple expression for sinusoidal correlation ToF (see \eqref{eq:atan-reconstruction}), we cannot easily invert our correlation function for our pulsed approach (cf.~\eqref{eq:correlation_function}). 
Instead we perform a calibration step in which we measure the raw fraction with a homogeneous calibration target (white diffuse plate) at a fixed distance to sensor and light source. 
Figure~\ref{fig:measurement_calibration} (left and middle) visualizes the calibration process for a single pixel. \textbf{First}, we perform measurements with $512$ equally spaced phase shifts $\doiphaseshift$. 
This reveals the upper and lower plateaus of the (ideally) rectangular correlation function. 
In theory, these plateaus have equal absolute value and hence the phase of maximum sensitivity is at $\rawfraction = 0$. 
However, for real measurements we have to compute a \textit{zero equivalent} value which relates to the actual phase of maximum sensitivity exactly between the plateaus. 
This value refers to the depth $\depth$ at which the calibration target 
is placed. We estimate the limits of the sensitive region $\sensitiveregion$. 
\textbf{Second}, we measure $\rawfraction$ within the sensitive region by stepping over the corresponding $\doiphaseshift$ with as high precision as possible (14 bit). 
This yields lookup values that could be used directly for estimating phases from measured values. However, noisy measurements introduce ambiguities into a numerical inversion, 
which requires a smoothing step to obtain a monotonously rising function for unambiguous phase estimation. To this end, we fit a univariate spline representation to the measurements on a per-pixel level. The resulting lookup table can 
now be used to estimate the phase offset from $\doiphase$, which in turn can be controlled to have a desired value. 
\textbf{Third}, we note that not necessarily all pixels of a C-ToF sensor exhibit the exact same behaviour, which in our case leads to a different spline representation and phase of maximum sensitivity per pixel. These values are distributed around the reference phase, which corresponds to our fixed distance. As we can only chose a single depth of interest $\doidepth$ and corresponding phase $\doiphase$, we obtain the zero equivalent for each single pixel and subtract it from the median over all pixels. This way we obtain a calibration mask that employs a per-pixel phase correction with respect to the DOI.
\paragraph{Validation.}
\begin{figure}
\begin{minipage}{0.49\linewidth}
 \includegraphics[width=\linewidth]{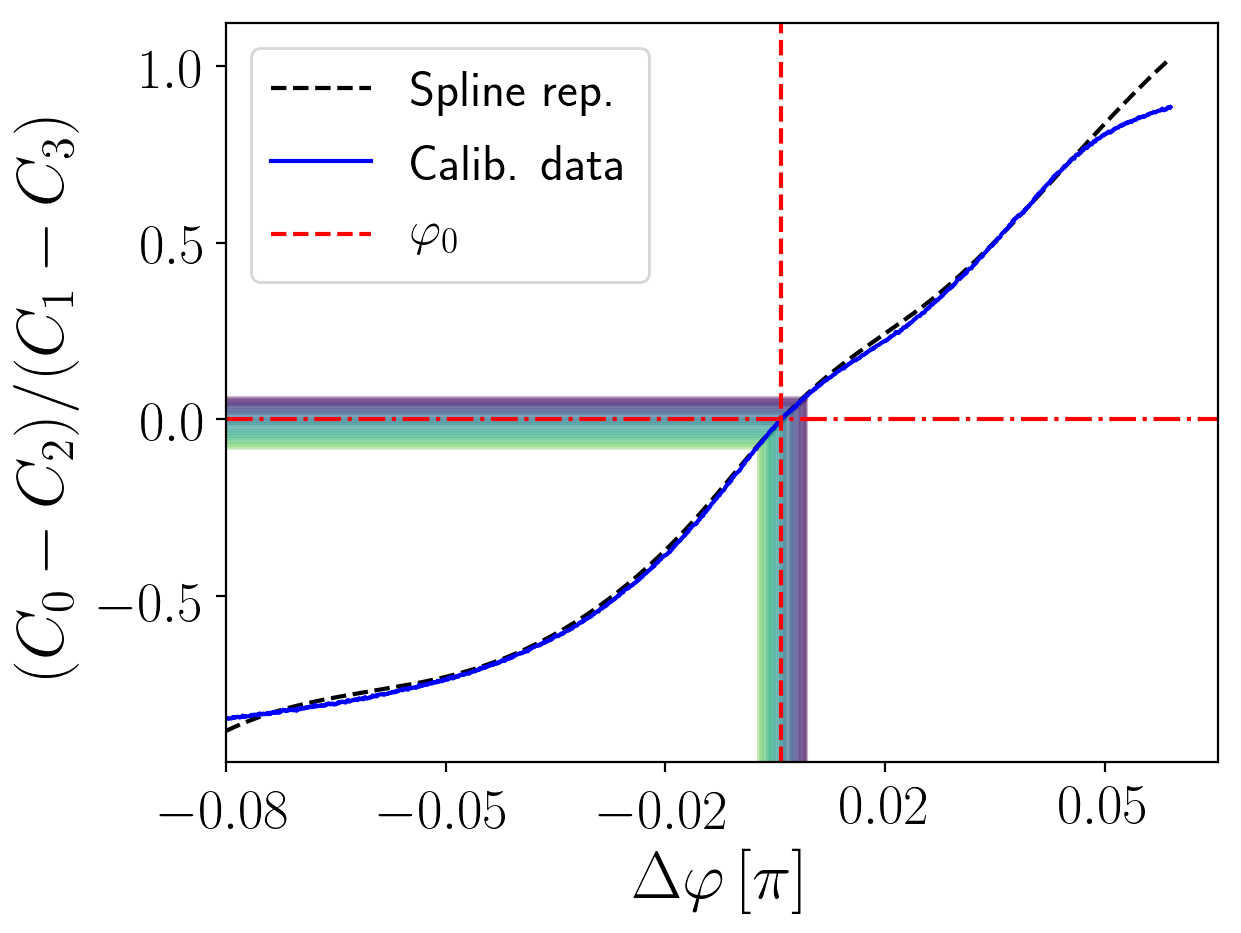}
\end{minipage}
\begin{minipage}{0.49\linewidth}
 \includegraphics[width=\linewidth]{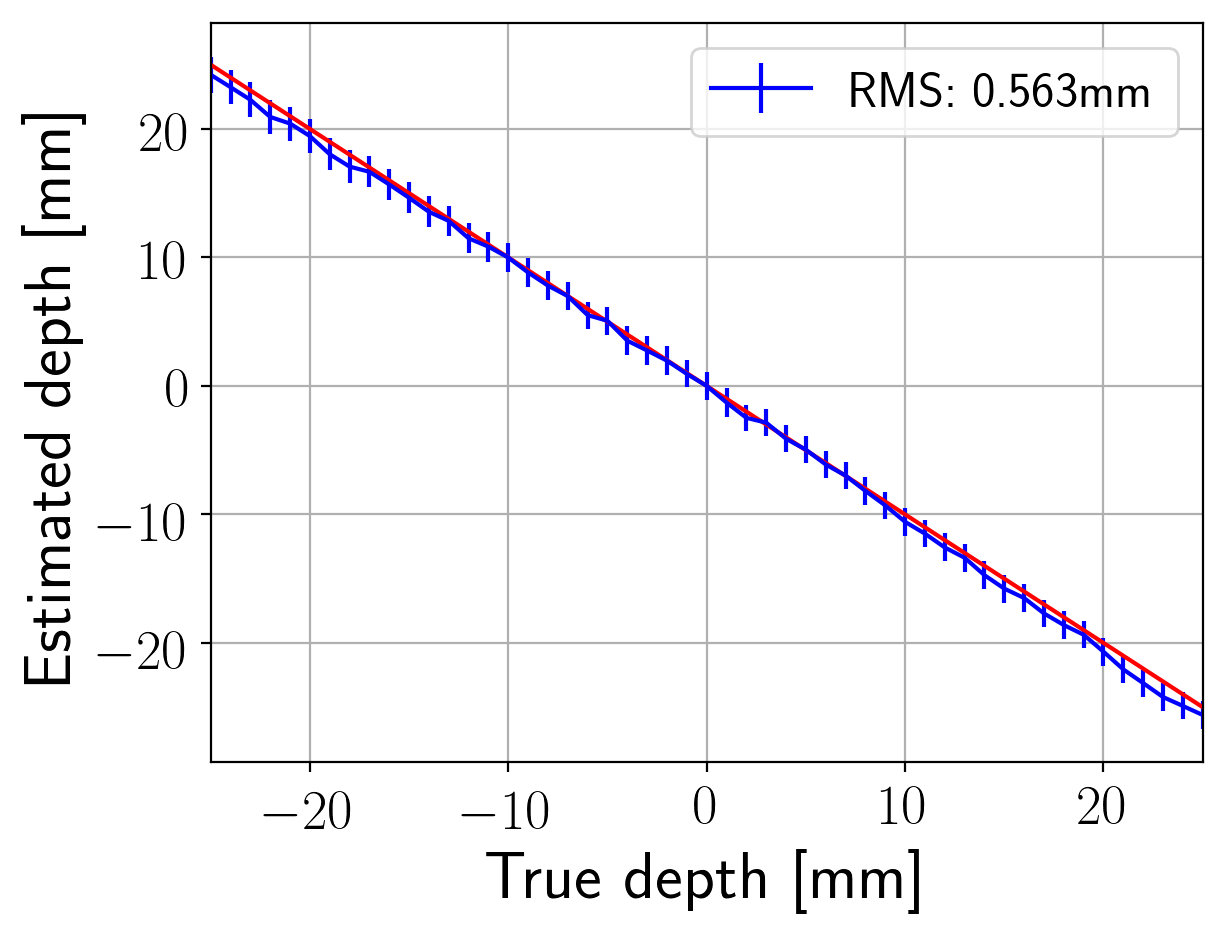}
\end{minipage}
 \caption{Example validation procedure for camera pixel $(60,80)$. 
 \textbf{Left}: The acquired calibration data and its fitted spline representation. The extracted phase of maximum sensitivity $\doiphase$ is given as a dashed red line. 
 To validate our calibration, we offset the mirror on the linear rail (cf.~\figref{fig:CToF-scheme}) by values within $[-$2.5$,$2.5$]$\,cm from the reference depth used for calibration. 
 We invert the phase value from the measurement value using the spline representation. The measurements and phase estimates are color coded from yellow to blue, representing the order of measurements. 
 \textbf{Right}: All 51 measurements, averaged over all camera pixels, plotted against the ground truth depth. The RMS error is approximately $0.6$\,mm.}
 \label{fig:validation_procedure}
\end{figure}
To validate our calibration, we need to assess how closely we can reconstruct changes of depth within a scene with the available calibration. As the calibration allows to measure an offset from $\doiphase$, we perform measurements with 
a planar calibration target again but now we change the distance the light has to travel by offsetting the mirror on the linear rail (cf.~\figref{fig:CToF-scheme}). 
We perform a total of 51 measurements with offsets in the range of $[-$2.5$,$2.5$]$\,cm at 1\,mm accuracy with respect to the reference depth used for calibration. 
The results (cf.~\figref{fig:validation_procedure}) indicate a good match between our depths obtained from a phase estimate using the spline representation for inversion and the ground truth depth values. 
Note that the validation is performed within a close range around the phase of maximum sensitivity, well within the sensitive range.
\begin{figure}
 \begin{minipage}{0.49\linewidth}
  \includegraphics[width=\linewidth]{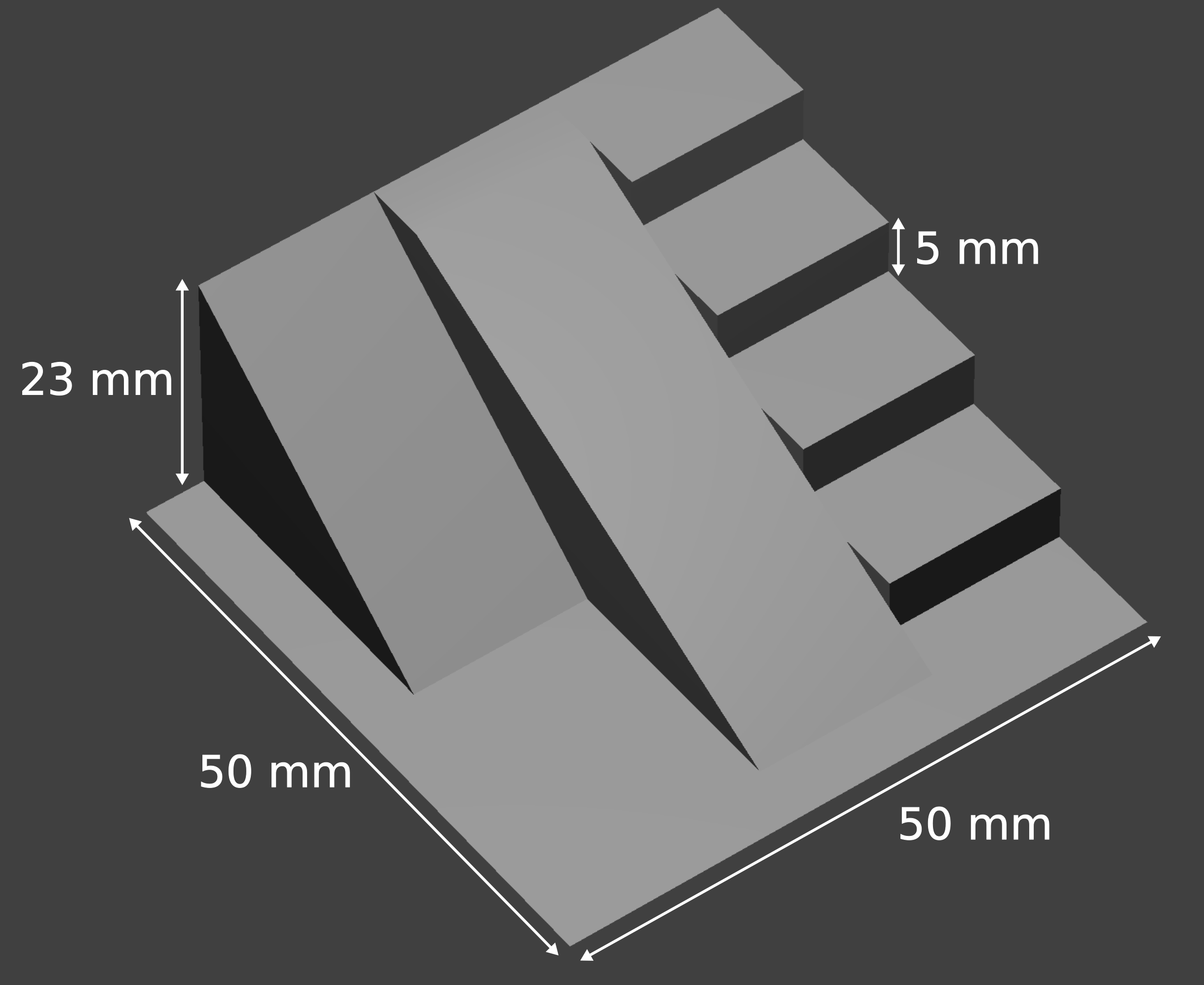}
 \end{minipage}
 \begin{minipage}{0.49\linewidth}
  \includegraphics[width=\linewidth]{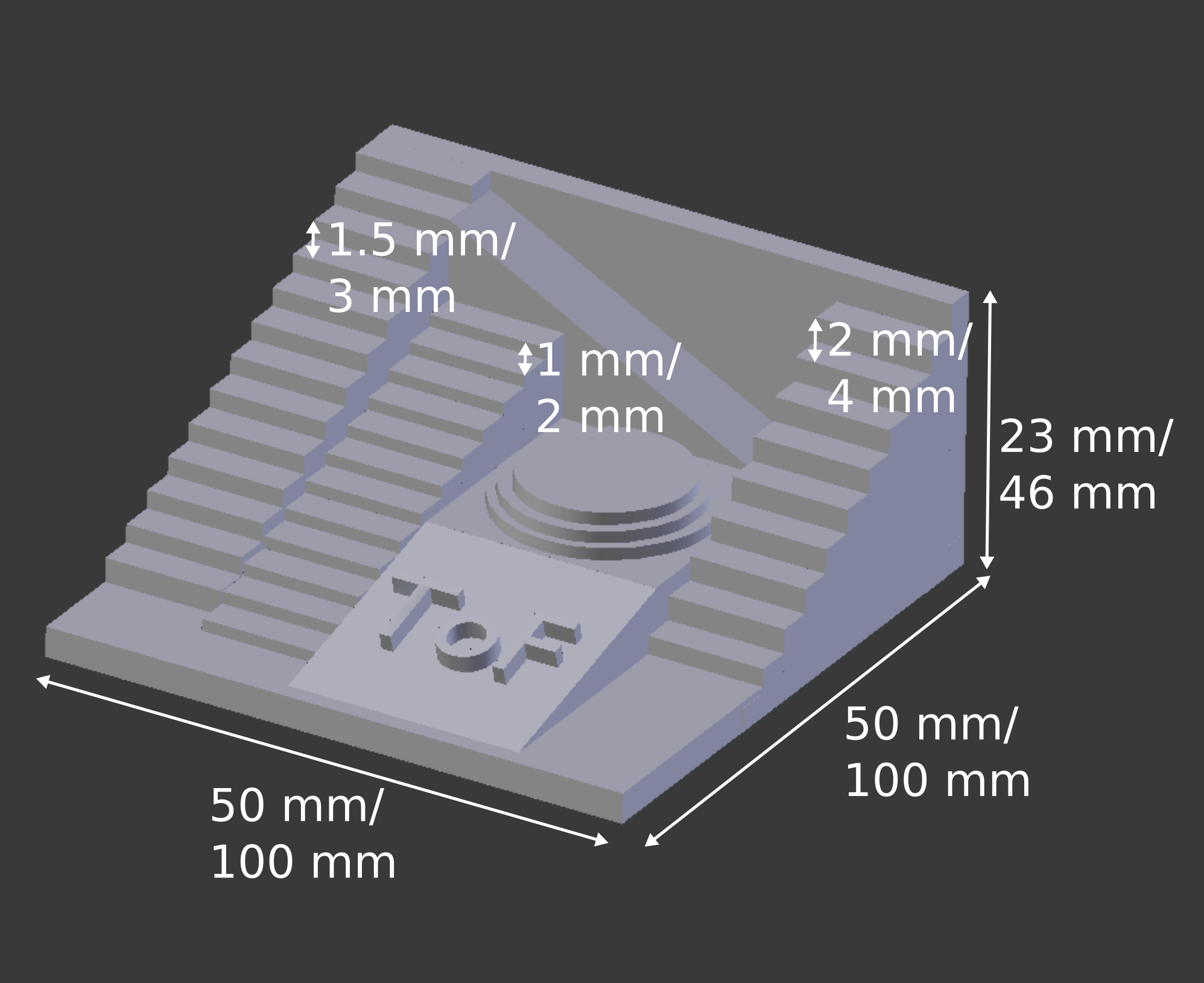}
 \end{minipage}
 \caption{Pictures of the 3D models we printed and measured using our pulsed ToF approach. 
 \textbf{Left}: Standard target box consisting of 2 ramps and stairs with a step height of 5\,mm. This scene is used to validate the working 
 principle of our approach.
 \textbf{Right}: Target box with detailed stairs. We generated 2 variants of this scene with the different measures separated as 1.5\,mm\,/\,3\,mm. 
 The most detailed stairs have a step height of 1\,mm and are considered the limit test case for our method. 
 }
 \label{fig:measurement_targets}
\end{figure}
\begin{figure*}[t]
\includegraphics[width=\linewidth]{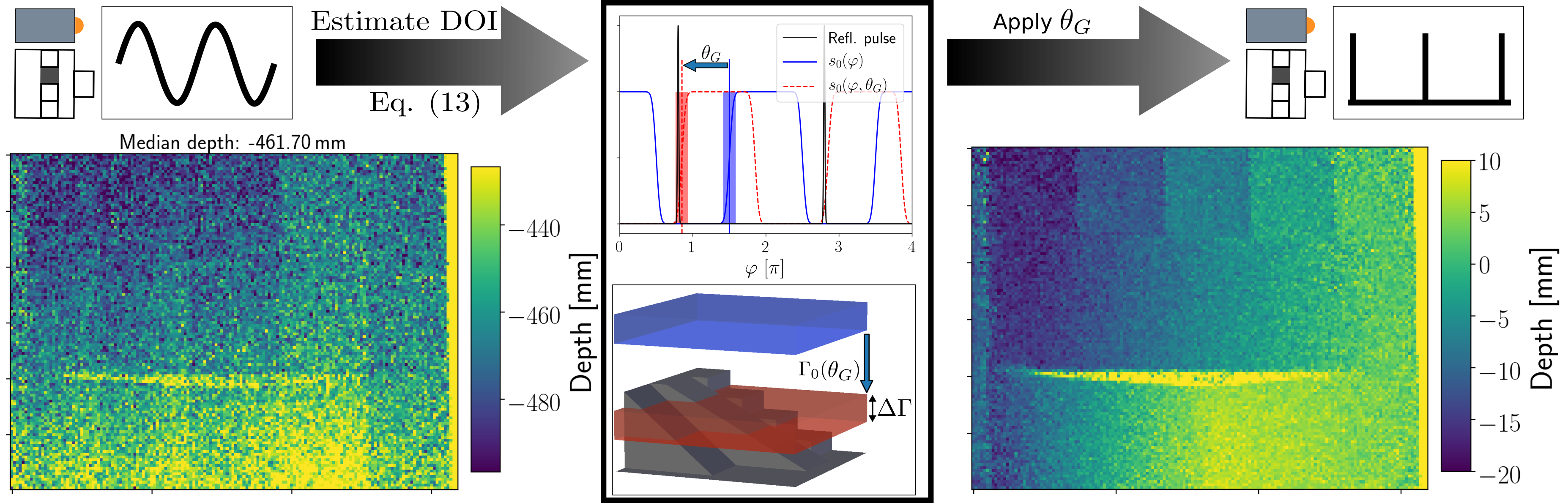}
\caption{
(\textbf{Left}): We perform a single ToF measurement with sinusoidal modulation and demodulation signals at 1\,mW -- The low SNR disallows to obtain a high 
resolution depth map but instead returns a rough depth estimate (about 50\,cm) for features of the scene we are interested in, the depth of interest (DOI). 
(\textbf{Middle}): Using~\eqref{eq:doi} we compute the phase shift required to let the reflected pulse and the signal edge coincide, focussing our 
measurement onto the DOI $\doidepth$. The shaded rectangles here illustrate the sensitive range surrounding the DOI, both in phase space and for 3D depths. 
(\textbf{Right}): We perform a single measurement at 1\,mW after applying the shift and switching the mode of operation to our pulsed acqusition. 
The obtained depth reconstruction exhibits well improved depth resolution for all scene features that lie within the sensitive range.
}
\label{fig:measurement_procedure}
\end{figure*}
\paragraph{Performing a PC-ToF measurement.}
After validation, our measurement procedure (cf.~\figref{fig:measurement_procedure}) is straightforward: We first acquire a rough depth estimate using a low power 
C-ToF measurement with sinusoidal modulation and demodulation with our system. With this, we obtain a rough estimate of the depth the object of interest is located at and adjust $\doiphaseshift$ such that the 
DOI is matched and interesting scene features lie within the sensitive range. Another measurement, now in pulsed operation delivers a much better 
resolved depth estimate. All measurements are obtained using the parameters given in \tabref{tab:hardware}, whereat the only difference between the operation modes is the usage of the different modulation signals. 
 \section{Results and conclusion}
\begin{figure}
\includegraphics[width=\linewidth]{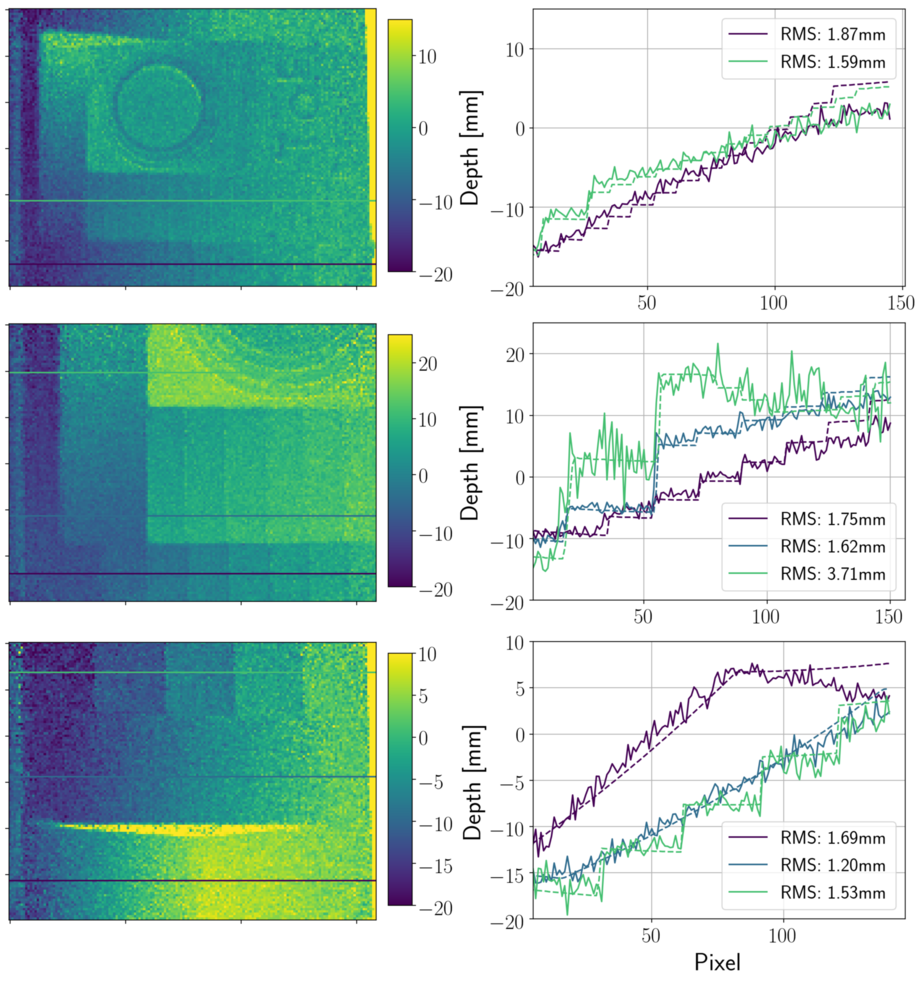}
\caption{
Depth maps (\textbf{left}) and comparison of depth slices (\textbf{right}) with our simulated ground truth data. All slices taken across stairs are 
averaged over 5 pixels in vertical direction. From top to bottom, the step heights of the stairs are 1\,mm (green), 1.5\,mm (violet); 2\,mm (blue), 3\,mm (violet); 5\,mm (green). The remaining three slices (middle, green; bottom, blue and violet) are taken across a more complex inhomogeneous geometry and two slopes with same maximum height but different length respectively.
}
 \label{fig:PToF-Results-1}
\end{figure}
\begin{figure}
\includegraphics[width=\linewidth]{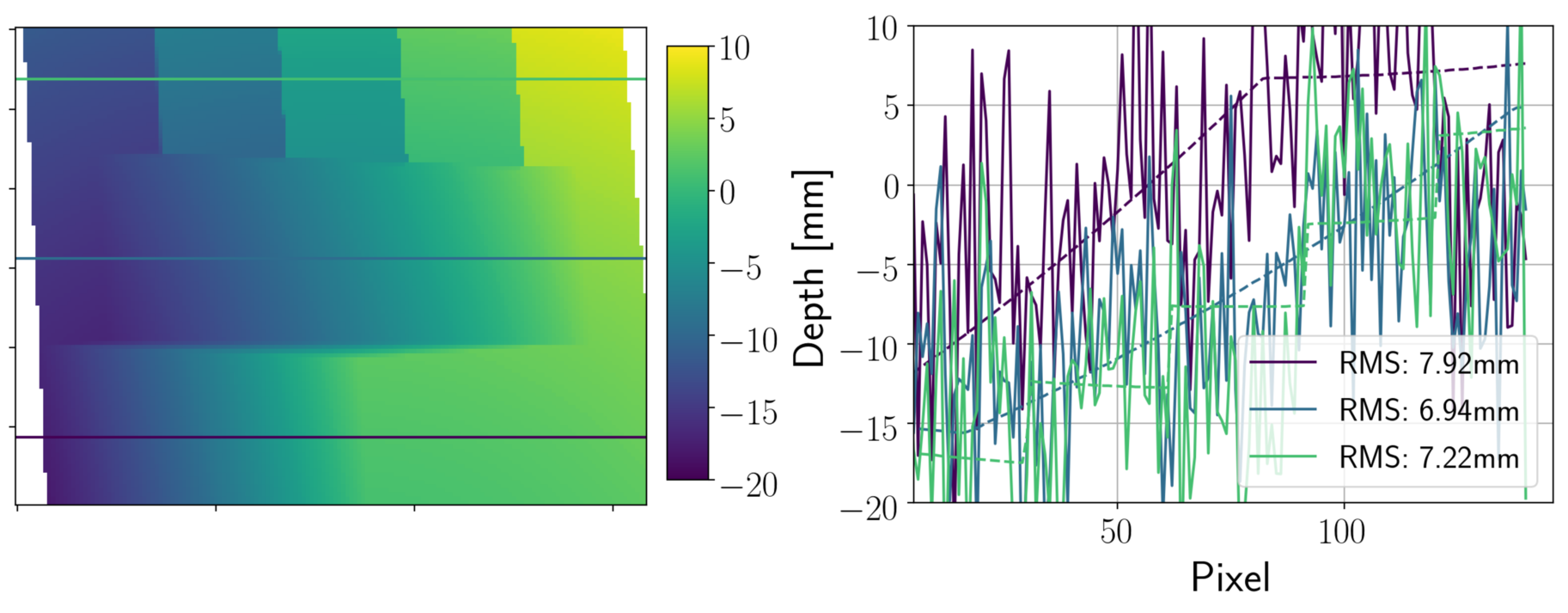}
 \caption{\textbf{Left}: Ground truth depth map obtained from our simulation for the first target. Note that our pose estimation does not deliver a perfect match, comparison with \figref{fig:PToF-Results-1} (lower left) reveals that the 
 orientation does not perfectly match, leading to systematic errors.
	  \textbf{Right}: Comparison of the C-ToF measurement with sinusoidal modulation of the scene at 1\,mW (cf.\figref{fig:measurement_procedure}) with the ground truth depth for the indicated slices. The measured depth overall follows the trend of the ground truth but is too noisy to reveal any details.
	  }
 \label{fig:gtdmap-stofslice}
\end{figure}
To assess the capabilities of our approach, we designed and 3D-printed three different targets, 
depicted in \figref{fig:measurement_targets}. The targets dimensions are chosen such that we cover a range of relative 
depth differences, starting from discrete steps of 5\,mm down to 1\,mm. Given the results from the 
validation measurement (cf.~\secref{sec:setup}), we regard signal changes that originate from such small depth differences 
as the limit our approach can resolve; in fact, we cannot assume to exactly match DOI and real depth (see \secref{sec:theory}), but as long as 
the depth remains in the sensitive range, the measurement is performed at high albeit not maximum sensitivity. \\
\figref{fig:PToF-Results-1} shows the results of our measurements for three distinct setups, each with a different target. 
All of our results show an increasing discrepancy between ground truth and measurement towards the right edge of the depth maps, best visible in the depth slices in~\figref{fig:PToF-Results-1}. 
There are two main factors from which this systematic error could result: 
First, we compute our ground truth via simulation, which relies on the correct pose estimation for the 3D models. This is a difficult task on its own, especially given the low resolution images. 
For example, \figref{fig:gtdmap-stofslice} reveals that indeed our pose estimation is not perfectly accurate, but a tilt is clearly visible. 
Second, our simulation assumes only a simple pinhole camera and the conjunction of light source and camera at the same position. However, in reality it is not possible to place both at the exact same location, an offset comes into play which will result in deviations with spatial dependency. 
In addition our approach, as well as all available correlation ToF systems, inherently suffers from the so-called multipath interference problem (MPI): 
Whenever multiple light paths from different scene points end up in one sensor pixel, the resulting depth estimate for this pixel is shifted to higher values, best visible near corners (\figref{fig:PToF-Results-1} bottom left). \\
Still, our approach reveals depth differences as fine as 2\,mm (cf.~\figref{fig:PToF-Results-1} middle), utilizing measurements with only 1\,mW average power and 10\,MHz modulation frequency by 
choosing a depth of interest based on a rough depth estimate. The approach is inherently dependent on the pulse shape of the modulation and rise time of the demodulation signal and hence the shape of 
those signals. The modulation frequency is only a secondary factor.\\
In future work, we would like to test the limits of the approach for more contemporary ToF sensors, which operate at frequencies of  100\,MHz or more. With higher frequencies, we expect shorter rise times of the sensor modulation and at the same time we can accumulate more laser pulses in the same exposure time, increasing the SNR. This should also allow us to test our approach on scenes with larger depth ranges - focusing on different targets that ideally are allowed to be meters apart and reconstructing the depths within a few centimeters around the respective DOI with high accuracy. 
Future work should also include a way to circumvent the calibration procedure described in~\secref{sec:setup}: 
Using approximations and by allowing more realistic modulation and demodulation signals (e.g., the saddlepoint found in the measurement of $\rawfraction$ on the rising signal edge - \figref{fig:validation_procedure}), 
we would like to find a closed-form solution to invert the measurement formalism. A calibration could then find the parameters of this analytic solution instead of generating a lookup table. \\
In the end, our approach is most suited for low power scenarios, such as mobile devices or for static measurement scenarios, where a detailed depth estimation for objects at a distinct range is required. With PC-ToF we trade the global sensitivity of a standard C-ToF setup (covering the unambiguity range) for highly increased sensitivity around a depth of interest: The broader the sensitive range, the less the maximum sensitivity and vice versa. In turn this allows for a task-specific tailoring of the pulse width and rise time, the two parameters that drive the depth resolution and sensitive range achievable with PC-ToF.

\FloatBarrier
\paragraph*{Acknowledgements}
This work was supported by the Computational Imaging Group from the Stuttgart Technology Center of Sony Europe B.V. and by the ERC Starting Grant ECHO.
{\small
\bibliographystyle{ieee}
\bibliography{egbib}
}

\end{document}